\documentclass[%
 amsmath,amssymb,
 aps,
 prl,
 longbibliography,
 lengthcheck,%
]{revtex4}

\usepackage{graphicx}
\usepackage{dcolumn}
\usepackage{epsfig}
\usepackage{bm}
\usepackage{hyperref}
\usepackage{hhline}
\begin{document}

\title{Generalized Parton Distributions from Deep Virtual Compton
Scattering at CLAS.}
\author{M. Guidal} 
\address{Institut de Physique Nucl\'eaire d'Orsay, 91405 Orsay, France}
\date{\today}

\begin{abstract}
We have analyzed the beam spin asymmetry and the longitudinally polarized
target spin asymmetry of the Deep Virtual Compton Scattering process,
recently measured by the Jefferson Lab CLAS collaboration. Our aim is to 
extract information about the Generalized Parton Distributions of the proton. 
By fitting these data, in a largely model-independent procedure, we are able
to extract numerical values for the two Compton Form Factors 
$H_{Im}$ and $\tilde{H}_{Im}$ with uncertainties, in average, of the order
of 30\%.

\end{abstract}


\maketitle

The study of Generalized Parton Distributions (GPDs) is currently one 
of the most intense fields of research in hadronic physics, theory-wise
as well as experiment-wise. GPDs give
access in an unprecedented way to part of the complex composite structure 
of the nucleon (or more generally of hadrons), which, until now, is not fully 
calculable from first principles of Quantum Chromo-Dynamics (QCD). 
For instance, nucleon GPDs encode, in a frame where the nucleon has a 
quasi-infinite momentum in a certain direction (the so-called 
``infinite momentum frame"), the longitudinal momentum distributions of the 
quarks and gluons in the nucleon, their transverse spatial distribution
and, overall, the correlation between these two latter distributions,
which is brand new information. As a consequence of this longitudinal
momentum-transverse space
correlations, there is the possibility to access the contribution
of quarks to the orbital momentum of the nucleon. This is of
great interest for the notorious ``spin puzzle" of the nucleon,
a long-standing issue in nucleon structure studies.
We refer the reader to Refs.~\cite{muller,ji,rady,collins,goeke,revdiehl,
revrady,barbara} for the original theoretical articles and recent comprehensive 
reviews on GPDs and for details on the theoretical formalism.

\begin{figure}[htb]
\epsfxsize=9.cm
\epsfysize=10.cm
\epsffile{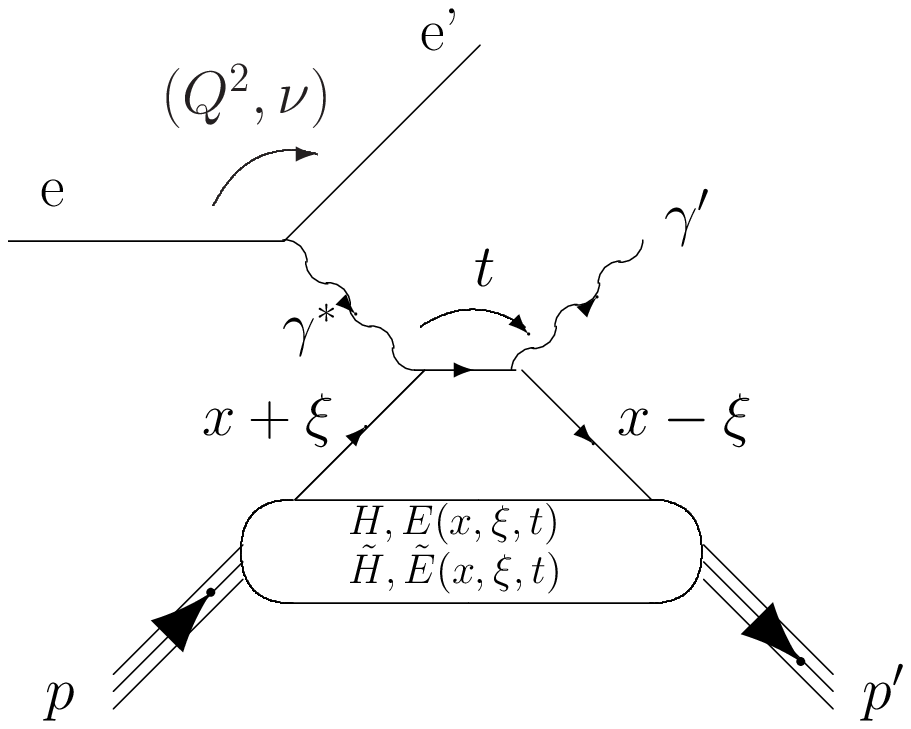}
\vspace{-4.2cm}
\caption{The handbag diagram for the DVCS process on the proton 
$ep\to e'p'\gamma '$. There is also a crossed diagram which is not shown here.}
\label{fig:dvcs}
\end{figure}

Nucleon GPDs are the structure functions which are accessed, through 
the factorisation property of QCD, in the hard exclusive
electroproduction of a meson or a photon off the nucleon. If we focus on 
quark GPDs, the golden channel to access them is the Deep Virtual Compton 
Scattering (DVCS) process, due to the purely electromagnetic nature of the 
perturbative part of the ``handbag" diagram. This latter diagram is 
schematized in Fig.~\ref{fig:dvcs}. At large $Q^2=(e'-e)^2$ and small $t=(p-p')^2$, 
this process in which the \underline{same} quark (or antiquark) absorbs the incoming 
virtual photon and radiates the final real photon, is predicted to be the
dominant one. The quantities $x+\xi$ and $x-\xi$ are the longitudinal momentum fractions 
of the initial and final quark (or antiquark) respectively, where
$\xi=\frac{x_B}{2-x_B}$ and $x_B=\frac{Q^2}{2m\nu}$ (with 
$\nu=E_{e^\prime}-E_e$) is the standard Deep Inelastic Scattering (DIS) 
variable.

For DVCS on the proton, several experimental observables measured in different 
kinematical regimes have been published this past decade~: cross sections
(beam-polarized and unpolarized) from the JLab Hall A 
collaboration~\cite{franck}, beam spin 
asymmetries (BSA) ~\cite{fx,stepan} and longitudinally polarized target
asymmetries (lTSA) ~\cite{chen} from the CLAS collaboration 
and a series of correlated beam-charge, beam-spin and transversely
polarized target spin asymmetries from the HERMES
collaboration~\cite{dvcshermes,ave,hermes}. 
The question arises: from this large harvest of experimental observables, 
with still much more to come, how can the GPD information be extracted ? 
The issue is not trivial as we recall that:
\begin{itemize}
\item the Bethe-Heitler (BH) process is another mechanism which 
leads to the same final state $ep\to ep\gamma$ as DVCS.
In the BH process, the final state photon is radiated by the incoming or 
scattered electron and not by the nucleon itself. Therefore, the BH process,
which dominates the cross sections in some kinematic regions,
carries no information about GPDs. However, it is relatively precisely 
calculable in Quantum Electro-Dynamics (QED) given the nucleon 
form factors, which are quite precisely known at the 
kinematics we are presently interested in, i.e. small $t$.
\item in the QCD leading twist and leading order approximation, which is the 
frame of this study, there are, for DVCS, four independent GPDs: $H, E, 
\tilde{H}$ and $\tilde{E}$ which correspond to the various spin and helicity 
orientations of the quark and nucleon in the handbag diagram.
These four GPDs depend on three variables $x$, $\xi$ and $t$. 
Decomposing the DVCS amplitude into real and imaginary parts leads to 
eight GPD-related quantities. We will call them the Compton Form Factors
(CFFs) and they are the quantities which can in principle be extracted 
from DVCS experiments. 
Following our conventions introduced in Refs.~\cite{fitmick,fithermes}, 
these eight CFFs are:
\begin{eqnarray}
&&H_{Re}=P \int_0^1 d x \left[ H(x, \xi, t) - H(-x, \xi, t) \right] C^+(x, \xi),\label{eq:eighta} 
\\
&&E_{Re}=P \int_0^1 d x \left[ E(x, \xi, t) - E(-x, \xi, t) \right] C^+(x, \xi),\label{eq:eightb} 
\\
&&\tilde{H}_{Re}=P \int_0^1 d x \left[ \tilde H(x, \xi, t) + \tilde H(-x, \xi, t) \right] C^-(x,
\xi),\label{eq:eightc} 
\\
&&\tilde{E}_{Re}=P \int_0^1 d x \left[ \tilde E(x, \xi, t) + \tilde E(-x, \xi, t) \right] C^-(x,
\xi),\label{eq:eightd} 
\\
&& H_{Im}=H(\xi , \xi, t) - H(- \xi, \xi, t),\label{eq:eighte} \\
&& E_{Im}=E(\xi , \xi, t) - E(- \xi, \xi, t),\label{eq:eightf} \\
&& \tilde{H}_{Im}=\tilde H(\xi , \xi, t) + \tilde H(- \xi, \xi, t) \;\;\;\;\text{and}\label{eq:eightg} \\
&& \tilde{E}_{Im}=\tilde E(\xi , \xi, t) + \tilde E(- \xi, \xi, t)\label{eq:eighth} 
\end{eqnarray}

with 
\begin{equation}
C^\pm(x, \xi) = \frac{1}{x - \xi} \pm \frac{1}{x + \xi}.
\end{equation}

\end{itemize}

In the QCD leading twist and leading order approximation, these eight CFFs depend 
only on $\xi$ (or equivalently $x_B$) and $t$. 

In Refs.~\cite{fitmick,fithermes}, we have developped and applied a largely 
model independent fitting procedure which, at a given experimental ($x_B$, $-t$) 
kinematic 
point, takes the CFFs as free parameters and extracts them from DVCS observables using the
well established QCD leading twist and leading order DVCS+BH theoretical amplitude. The 
expression of this amplitude can be found, for instance, in Ref.~\cite{vgg1}. With this 
procedure, we have fitted in Ref.~\cite{fitmick} the JLab Hall A proton DVCS beam-polarized 
and unpolarized cross sections. We could then extract the $H_{Im}$ and $H_{Re}$ CFFs at 
$<x_B>=0.36$ and for several $t$ values with average uncertainties of the 
order of 35\% for $H_{Im}$ and larger for $H_{Re}$. In Ref.~\cite{fithermes}, we have 
fitted several HERMES beam-charge, beam-spin and transversely polarized target 
spin asymmetries. We could then extract at $<x_B>=0.09$ and for several $t$ values, the same CFFs $H_{Im}$ and 
$H_{Re}$ with roughly similar uncertainties as for JLab. 

The sources of uncertainty in our approach stem, on the one hand, 
from the experimental errors of the data that we fit, and on the other hand,
from the fact that we take in our fits practically all CFFs as free
parameters, with relatively large and conservative bounds. There are 
therefore minimum conjectures and surmises in our work, which is 
certainly highly valuable. However, given that we generally fit limited sets 
of data and observables, our problem is in principle underconstrained. The 
consequence is that there are maximum correlations and interferences between 
our fitted parameters, hence the relatively important error bars in our results. 
In these extremely conservative conditions, it is nevertheless remarkable that we 
managed, in our previous works, to extract several CFFs, at different energies, with 
well-defined uncertainties, fitting the very limited available data. The reason for this
convergence of a few CFFs,in spite of the underconstrained nature of the problem,
 is that some observables are in general dominated by 
some particular CFFs, like BSAs by $H_{Im}$ and beam charge asymmetries and 
cross-sections by $H_{Re}$. Our uncertainties 
can only decrease in the future as, on the one hand, larger (and more precise) 
sets of data and observables sensitive to different CFFs become available and, 
on the other hand, theoretical constraints which allow to reduce, in the most 
model-independent way possible, the range of variation, or even the number, of the 
CFFs come forth (for instance, dispersion 
relations~\cite{dis1,dis2,dis3,dis4}). 

We mention that related DVCS fitting studies have been published this past 
year~\cite{rv,kumer}. They resulted also in the extractions of the $H_{Im}$ 
and $H_{Re}$ CFFs with central values consistent with ours, although
with smaller uncertainties. These fits are however model dependent.
They either neglected all CFFs other than $H_{Im}$ and $H_{Re}$ or (and)
assumed a functional shape for the CFFs, allowing to fit several
($x_B$, $-t$) points simultaneously. The uncertainty associated
to the model dependence and hypothesis entering these approaches
is then very difficult to estimate and to take into account properly. 
Nevertheless, each of these model-dependent and -independent approaches
have their own merits and values. The fact that they all 
result in consistent and compatible central values for the fitted $H_{Im}$ 
and $H_{Re}$ CFFs gives mutual support and credit for each of them.

In this letter, we continue our model-independent fitting approach focusing 
this time on the CLAS BSAs and lTSAs, which we had not considered so far.
As was already mentioned in our pioneer work on the subject~\cite{fitmick}, 
fitting only the CLAS BSAs, without any model-dependent hypothesis
or input, was not constraining enough. In other words, fitting only BSAs, 
i.e. only one observable,
with seven or eight unconstrained CFFs does not lead to well-defined solutions. 
However, inspired by our recent rather succesful fit of a series of 
HERMES asymmetries~\cite{fithermes}, we now want to investigate if,
with the addition of a new observable, namely the lTSAs, to be
fitted simultaneously with the BSAs, progress can be made. Indeed, lTSAs
have been measured by the CLAS collaboration as well and have 
actually received little attention from GPD phenomenologists so far. The BSAs being 
in general dominantly sensitive to $H_{Im}$ and the lTSAs to 
$\tilde{H}_{Im}$~\cite{fitmick,kirch}, our expectation is to extract
some quantitative information on these two particular CFFs, which would be
brand new information for $\tilde{H}_{Im}$ in particular. 

Let us describe these CLAS data. Regarding BSAs, the Hall B collaboration 
has measured their $\phi$ distribution at 57 ($x_B$, $-t$, $Q^2$) points ($\phi$ 
is the standard angle between the leptonic and hadronic planes of the DVCS process). 
The values of the $x_B$ variable extend from $\approx$0.13 up to $\approx$0.46, those
of the $-t$ variable from $\approx$0.13 GeV$^2$ up to 
$\approx$1.3 GeV$^2$ and those of the $Q^2$ variable from $\approx$1.2 GeV$^2$ up 
to $\approx$3.3 GeV$^2$. The amplitude
of these BSAs range from $\approx$ 0 up to $\approx$0.3~\cite{fx}.

Regarding lTSAs, the data is much more scarce: in Ref.~\cite{chen}, only 
their $\sin(\phi)$ moment is available at 
a few ($x_B$, $-t$, $Q^2$) points. The average kinematics of this
whole set of data is $<\xi>$=0.16 (i.e. $<x_B>\approx$0.275), $<-t>$=0.31 GeV$^2$
and $<Q^2>$=1.82 GeV$^2$. Within this phase space, the lTSA $\sin(\phi)$ moments
(which we will designate as $A_{UL}^{\sin\phi}$) following, for instance, 
Ref.~\cite{fithermes}) have been extracted differentially, either for three $x_B$ 
values ($\approx$0.20,
0.29 and 0.40) or for three $-t$ values ($\approx$0.15, 0.24 and 0.43 GeV$^2$).
There are therefore six lTSAs available, which are actually
not statistically independent since they are issued from the same
set of data which has been binned either in $x_B$ or in $t$. However, even if scarce,
these data are extremely valuable as we will show in the following. Their 
amplitudes range from $\approx$ 0.07 to $\approx$ 0.38,
with uncertainties extending from 30\% to more than 100\%.

Following our notation of Ref.~\cite{fithermes}, the BSAs can also
be denoted as $A_{LU}$ and the lTSAs moments as $A_{LU}^{\sin\phi}$.
We will use this notation on our figures.

In the first stage, we will see what information one can extract from
the simultaneous fit of one lTSA and one BSA whose ($x_B$, $-t$, $Q^2$)  
values approximatively match. In the second stage, we will fit simultaneously
one lTSA and the two or three BSAs which have the same ($x_B$, $-t$) values,
irrespective of the $Q^2$ value. Assuming the dominance
of the QCD leading twist and leading order DVCS contribution, i.e. that CFFs do not 
depend on $Q^2$, the idea is to improve the statistical accuracy and 
increase the constraints on our fitting procedure.

Among the six available lTSAs, we first focus on the lTSA which has been measured 
at the kinematic point ($x_B$, $-t$, $Q^2$) =(0.29,0.31,1.82).
Unfortunately, the BSAs and the lTSAs are issued from two different Hall B 
experiments so that they have not been measured at exactly the same average 
kinematics. The matching of the kinematics between the different observables
can thus be only approximate.  Among the 57 BSAs,
the two BSAs whose kinematics is the closest of the lTSA kinematics that
we focus on, are at (0.25,0.28,1.69) 
and (0.25,0.28,1.95). We therefore note at this stage the differences 
between the $x_B$ values (0.25 vs 0.29), the $-t$ values (0.28 vs 0.31)
and $Q^2$ values (1.69 or 1.95 vs 1.82). There is a third BSA which has the same
$x_B$ and $t$ values as the two BSAs just mentioned but whose $Q^2$ is equal 
to 2.21 GeV$^2$. We will use this extra BSA in our $Q^2$-independent ``second stage" 
fitting. We recapitulate these four kinematic points on which we presently focus
in Table~\ref{tab:kin}.

\begin{table}[h]	
\begin{tabular}{||c||c|c|c||}
\hline
      & $<x_B>$& $<-t>$& $<Q^2>$ \\ \hline
lTSA  & 0.29 & 0.31 & 1.82 \\ \hline
BSA1  & 0.25 & 0.28 & 1.69 \\ \hline
BSA2  & 0.25 & 0.28 & 1.95 \\ \hline
BSA3  & 0.25 & 0.28 & 2.21 \\ 
\hline
\end{tabular}	
\caption{Summary of the four kinematical points which have approximately the 
same ($x_B$, $-t$) values. We will fit four ``topologies": (lTSA+BSA1), (lTSA+BSA2),
(lTSA+BSA3) and (lTSA+BSA1+BSA2+BSA3).} 
\label{tab:kin}
\end{table}	

For the fitting procedure, as in Refs~\cite{fitmick,fithermes}, we minimize our 
theoretical calculations of the DVCS observables based on the well-known 
QCD leading twist and leading order DVCS+BH amplitude, 
by the standard $\chi^2$ function, using MINUIT~\cite{james}. We recall
that the parameters to be fitted are the CFFs of Eqs~\ref{eq:eighta}-\ref{eq:eighth}. 
As in Refs~\cite{fitmick,fithermes}, we have actually considered only seven CFFs,
setting $\tilde{E}_{Im}$ to zero. This is based on the theoretical guidance which approximates 
the $\tilde{E}$ GPD by the pion exchange in the $t$-channel whose amplitude is real. 
With the hypothesis of the dominance of the leading twist amplitude of the DVCS 
process, this is the only model-dependent assumption that enters our fitting procedure.
A last feature entering our fitting process is that we have to bound
the domain of variation of the fitting parameters. Without bounds, our fits
which are in general underconstrained, would probably not converge and/or
would yield values for the fitted parameters with infinite uncertainties.
Following what we have done and explained in details in Refs.~\cite{fitmick,fithermes},
we bound the allowed range of variation of the CFFs to $\pm$5 times some ``reference" 
VGG CFFs. VGG~\cite{vgg1,gprv} is a well-known and widely used model
which provides an acceptable first approximation of the CFFs, as shown
in our previous studies~\cite{fitmick,fithermes} and as will be confirmed
furtherdown in the present work.
We do not really consider this as a model-dependent input since this allowed
deviation of a factor $\pm$5 with respect to the VGG model values is extremely
conservative. We recall that GPDs have to satisfy a certain number of
normalization constraints in general, these being all fulfilled by the VGG model.
Finally, the problem at stake being non-linear and the parameters being correlated, 
we use MINOS for the uncertainty calculation on the resulting fitted 
parameters~\cite{james}.

Before presenting our results, we also want to outline the point that our aim
is to fit the lTSA and BSA(s) with the $\underline{same}$ CFFs, which therefore
should correspond to unique ($x_B$, $-t$) values. As the data of Table~\ref{tab:kin}
do not, unfortunately, have exactly the same kinematics, as mentioned previously,
there is an ambiguity
in defining the precise ($x_B$, $-t$) values of the fitted CFFs. We will consider that 
the CFFs that we will fit to the kinematic points of Table~\ref{tab:kin} correspond 
to the values of the BSA kinematics, i.e. (0.25, 0.28), as the BSA observable is 
in general the most significant statistically. There is clearly an approximation here, 
which we will make for the moment, lacking better solution, in order to make progress. 
This approximation is to some extent supported by the VGG model which predicts 
about 8\% difference for $H_{Im}$ between $x_B$=0.25 and 0.29 (at $-t$=0.28 GeV$^2$) 
and less than 2\% for $\tilde{H}_{Im}$ for 
this same kinematic change. We stress that this kinematical matching problem is of
a rather trivial nature and it is sufficient that future experiments, measuring
different observables, simply agree to analyze data at the same central kinematics
to avoid this extrapolation issue. 

\begin{figure}[htb]
\epsfxsize=9.cm
\epsfysize=10.cm
\epsffile{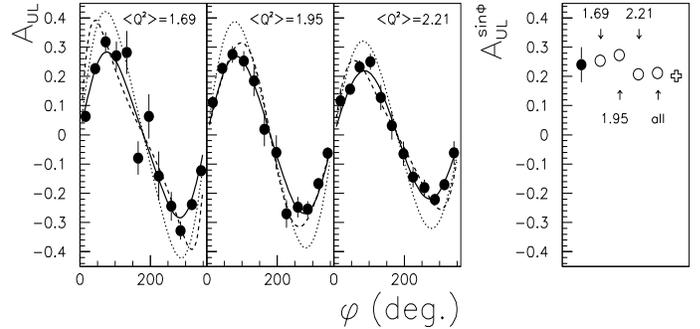}
\vspace{-4.2cm}
\caption{Comparison of our fit results with the experimental data.
The three left panels show the three experimental BSAs (i.e. $A_{LU}$)
as measured in Ref.~\cite{fx} (solid circles). The right panel shows the 
experimental lTSA moment (i.e. $A_{LU}^{\sin\phi}$)as measured in 
Ref.~\cite{chen} (solid circle). The four panels correspond to the 
four ($x_B$, $-t$, $Q^2$) kinematic points presented in Table~\ref{tab:kin}
(from left to right: BSA1, BSA2, BSA3 and lTSA). 
All four observables have approximately the same 
($x_B$, $-t$) values, taken as (0.25,0.28), but different $Q^2$ values.
On the BSA panels, the dashed solid curve is the result of our fit, fitting 
only the \underline{individual} BSA of the relevant panel with the lTSA, i.e.
from left to right, fit of (BSA1+lTSA), (BSA2+lTSA) and (BSA3+lTSA). The solid line
is the result of our fit, fitting \underline{simultaneously} the three
BSAs and the lTSA, i.e. (BSA1+BSA2+BSA3+lTSA). This latter fit therefore 
assumes that CFFs do not depend on $Q^2$. The fit results of these four 
``topologies" for the lTSA are displayed in the right panel.
On the BSA panels, the dotted curve is the prediction of the standard VGG 
model. Its prediction for the lTSA is displayed as the empty cross.}
\label{fig:data}
\end{figure}

We now display in Fig.~\ref{fig:data} the result of our fits. The left panel
shows the $\phi$ distribution of the three BSAs mentionned above and the right
panel the $\sin\phi$ moment of the lTSA. The dashed curves 
are the results of the fit of the lTSA (of the right panel) with each 
\underline{individual} BSA. The thick solid curves are the result of the 
fit of this same lTSA with the three BSAs 
\underline{simultaneously} (these three BSAs having the same ($x_B$, $-t$) values 
but different $Q^2$ values, see Table~\ref{tab:kin}). 
On the right panel, the four empty circles show the corresponding results of
the fit for the lTSA: the first three for the fit with the individual BSAs and 
the fourth one for the fit with the three
BSAs simultaneously. For comparison, we also show in this figure the predicted
results for the BSAs and the lTSA of the standard VGG model~\cite{vgg1,gprv}. 
It is seen that the VGG model overestimates the three BSAs by approximately 
0.1 (i.e. $\approx$ 30\%) and underestimates the lTSA by roughly the same proportion. 

\begin{figure}[htb]
\epsfxsize=9.cm
\epsfysize=10.cm
\epsffile{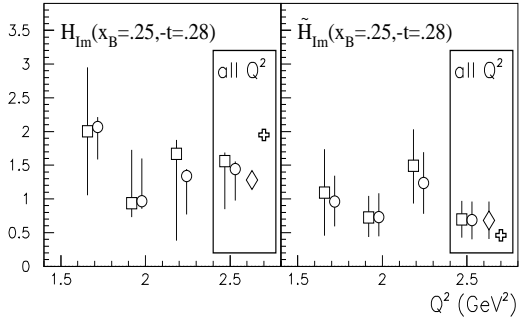}
\vspace{-4.2cm}
\caption{Results of our fits to the data displayed
in Fig.~\ref{fig:res} for the $H_{Im}$ and $\tilde{H}_{Im}$ CFFs.
The empty squares (circles) show our results when the boundary values 
of the domain over which the CFFs are allowed to vary is 5 (3)
times the VGG reference values. Both results have been slightly shifted 
left (right) from the central $Q^2$ value for sake of visibility.
The four sets of results correspond,
from left to right, to the fits (lTSA+BSA1), (lTSA+BSA2), (lTSA+BSA3) 
and (lTSA+BSA1+BSA2+BSA3) as indicated in Table~\ref{tab:kin}.
In particular, the (lTSA+BSA1+BSA2+BSA3) fit, which corresponds to an average $Q^2$
value of 1.95 GeV$^2$, is displayed within a box. The empty diamond
indicates the results of our fits, in the (lTSA+BSA1+BSA2+BSA3) 
``topology", when only the $H$ and $\tilde{H}$ GPDs are taken
as fitting parameters. The empty cross indicates the VGG prediction.}
\label{fig:res}
\end{figure}

We now show in Fig.~\ref{fig:res} the fitted values, with their error bars, 
of the only two CFFs, $H_{Im}$ and $\tilde{H}_{Im}$, out of seven,
that came out of our fitting procedure with finite MINOS uncertainties. 
We recall that the MINOS uncertainties correspond
to a deviation of 1 from the value which minimizes $\chi^2$. These
uncertainties can be asymmetric if the $\chi^2$ function is not symmetric
around the minimum, which is the sign of a non-linear problem in general.
The fact that only $H_{Im}$ and $\tilde{H}_{Im}$ converge in our 
fitting process reflects, as was mentioned earlier, the particular 
sensitivity of the BSA and lTSA observables, respectively, to these two CFFs.
The other five CFFs did not converge in our fitting procedure to some well 
defined value or domain: either their central value reached the boundaries of the allowed 
domain of variation or MINOS could not reach the $\chi^2$+1 value to
fully determine the associated uncertainties. These
features were well studied~\cite{fitmick,fithermes} in our earlier works. They 
reflect the fact that the contribution to the $\chi^2$ of those CFFs
which didnotconverge is relatively weak and that the fit is barely sensitive to them. However, it is
important to include them in our fit because they play a role, through
correlations, in the determination of the error bars on the two ``convergent"
CFFs.

In Fig.~\ref{fig:res}, we display four sets of results for
$H_{Im}$ and $\tilde{H}_{Im}$, which correspond to the four ``topologies" that
we mentionned earlier: three sets correspond to the fits
of the lTSA with each one of the BSAs at $Q^2$=1.69, 1.95 and 2.21 GeV$^2$
and the fourth set (in the box in Fig.~\ref{fig:res}) corresponds to the
simultaneous fit of the lTSA with the \underline{three} BSAs.
In this latter case, the underlying assumption is that CFFs do not depend 
on $Q^2$. As could be expected,
the resulting uncertainties are smaller for the both CFFs in this 
latter configuration, as more statistics and contraints enter into play.
We observe that all four configurations yield compatible results
within error bars, which are between 25\% and 50\% in average.
The simultaneous
fit of the three BSAs and of the lTSA yields an approximate
average of the fits using only one BSA and the lTSA. From the
uncertainties on the CFFs that we obtain, it is clear that no
QCD evolution or twist effect can be discerned. It then seems reasonnable
to fit simultaneously observables at the (approximately) same
($x_B$, $-t$) points and different $Q^2$ values.

In Fig.~\ref{fig:res}, we have also displayed, for each of the four fit topologies,
two results, aimed at illustrating the dependence of our results
on the boundaries of the domain of variation allowed for the CFFs. The empty squares 
show our results when the CFFs are limited to vary within $\pm$5 times the 
VGG reference values while the empty circles shows these results for boundaries
equal to $\pm$3 times these same VGG reference values. Of course, the smaller
the domain of variation, the smaller the uncertainties on the fitted CFFs. 
This shows the overall stability and robustness of our fitting 
process since the values of these boundaries do not affect strongly
the central values of the fitted CFFs. We also checked that
the fit results were not dependent on the precise starting values of the
CFFs when we begin our fit: irrespective of the
starting values, the minimization would essentially always converge
to the same central values and uncertainties for $H_{Im}$ and $\tilde{H}_{Im}$.

We further show in Fig.~\ref{fig:res} the result of our fit if we
take as fitting parameters only the $H$ and $\tilde{H}$ GPDs, i.e. four CFFs
($H_{Re}$, $\tilde{H}_{Re}$, $H_{Im}$ and $\tilde{H}_{Im}$), instead of seven.
For this configuration, we have fitted the three BSAs and the lTSA
simultaneously. The central values for $H_{Im}$ and $\tilde{H}_{Im}$
are in very good agreement with 
the ones previously determined when all CFFs were taken into account
(with, though, a some slight decrease of the central value of $H_{Im}$).
The obvious difference is that, as could be expected, the associated 
uncertainties are smaller, particularly for $H_{Im}$. There is not too much 
effect for 
$\tilde{H}_{Im}$. This can probably be attributed to the fact that when only
two GPDs enter the fit, the main source of uncertainty comes from the 
statistics of the observables to be fitted and no more from the correlations
between the fitting parameters. Indeed, $H_{Im}$ is mostly sensitive to the three BSAs
(which are simultaneously fitted) with each having smaller errors than
the lTSA, while $\tilde{H}_{Im}$ is mostly sensitive to the lTSA which has
a $\approx$ 25\% error bar. We do not display the comparison
of these ``2 CFFs" fit with the data in Fig.~\ref{fig:data} in order not to
overload the figure, but the $\chi^2$ is equally good to the fits
with all CFFs. In the latter case, the normalized $\chi^2$ is found to be 
equal to 1.27 while in the former case the normalized $\chi^2$
is 1.16. These good results obtained when fitting with only 
$H_{Im}$ and $\tilde{H}_{Im}$ mean that it is indeed possible to correctly 
fit the data with only these two GPDs instead of four. This however does not 
mean that this is the true solution and that the other GPDs should 
consequently be ignored or neglected. The large error bars that we obtain 
when all GPDs precisely reflect this lack of knowledge on the
other CFFs: our uncertainties incorporate all our ignorance about the 
other GPDs and all their full potential influence. 

We finally display in Fig.~\ref{fig:res} the predicted values of the corresponding 
VGG CFFs (empty crosses), which are $Q^2$ independent. 
It is noted that the VGG $H_{Im}$
tends to lie above the fitted $H_{Im}$ while the VGG $\tilde{H}_{Im}$
tends to lie below the fitted $\tilde{H}_{Im}$. This is a straightforward reflection
of what was observed in Fig.~\ref{fig:data} where the VGG BSAs curves 
were overestimating the data while the VGG lTSA point was underestimating
the data. The overestimation of the VGG $H_{Im}$, with respect to the 
fitted central value, was also observed in our study of the HERMES 
data~\cite{fithermes}.

We have so far focused on the particular lTSA measured at $x_B$=0.29
in order to establish and understand the basic features and
results of our approach. We now turn to the lTSA measured
at another $x_B$ value, i.e. $x_B$=0.40, for which there
are several BSAs which have neighboring ($x_B$, $-t$) values. These data 
points are indicated in Table~\ref{tab:kin2}. This time, none of the 
$Q^2$ values match each other and there is also a more significant difference 
between the $x_B$ values. Supported by our previous study which 
showed that the simultaneous fits of several observables at different
$Q^2$'s appeared to converge to some average of individual $Q^2$ fits, we 
attempt to fit simultaneously the four data points (i.e. 3 BSAs and 
1 lTSA) of Table~\ref{tab:kin2}. We mention the VGG predictions: 13\% change
for $H_{Im}$ between $x_B$=0.34 and $x_B$=0.40 (for $-t$=0.30 GeV$^2$)
and 7\% change for $\tilde{H}_{Im}$ for the same kinematical variation.  
In front of the anticipated $\approx$30\% error bars to be issued
from our fits, it is not unreasonnable to neglect, in a first approach, this small
$x_B$ variation. 

\begin{table}[h]	
\begin{tabular}{||c||c|c|c||}
\hline
      & $<x_B>$& $<-t>$& $<Q^2>$ \\ \hline
lTSA  & 0.40 & 0.31 & 1.82 \\ \hline
BSA1  & 0.34 & 0.30 & 2.3 \\ \hline
BSA2  & 0.34 & 0.28 & 2.63 \\ \hline
BSA3  & 0.35 & 0.29 & 2.97 \\ 
\hline
\end{tabular}	
\caption{Selection of kinematic points measured by the CLAS 
collaboration which have approximatively the same ($x_B$, $-t$)
values, around $x_B$=0.35.} 
\label{tab:kin2}
\end{table}	

Within this approximation, we are again able to 
extract values for the two CFFs $H_{Im}$ and $\tilde{H}_{Im}$ which we then
consider to 
correspond to the BSAs' kinematics ($x_B$, $-t$)=(0.35, 0.29), based on the 
statistical dominance of the BSAs. We show the resulting values 
of $H_{Im}$ and $\tilde{H}_{Im}$ in Table~\ref{tab:res}, 
along with the values we obtained previously for these two CFFs at $x_B$=0.25
when we fitted simultaneously all points of Table~\ref{tab:kin} (i.e. 
values of the data points in the ``box" of Fig.~\ref{fig:res}). 
We observe, although error bars are not negligible, the general trend that, 
at fixed $t$ ($\approx$0.28 GeV$^2$), $H_{Im}$ tends to increase,
as $x_B$ goes from 0.35 to 0.25, while $\tilde{H}_{Im}$ remains 
rather constant.

\begin{table}[h]	
\begin{tabular}{||c||c|c|c||}
\hline
$<x_B>$& $H_{Im}$& $\tilde{H}_{Im}$ \\ \hline
0.25 & $1.56^{+0.12}_{-0.71}$ & $0.69^{0.27}_{-0.27}$ \\ \hline
0.35 & $0.62^{+0.56}_{-0.18}$ & $0.63^{+0.60}_{-0.32}$ \\ \hline
\hline
\end{tabular}	
\caption{Results of our fits for the $H_{Im}$ and $\tilde{H}_{Im}$
CFFs from the CLAS BSAs and lTSA, at fixed $t\approx$0.28 GeV$^2$,
for two different $x_B$ values.} 
\label{tab:res}
\end{table}	

\begin{figure}[htb]
\epsfxsize=8.cm
\epsfysize=8.cm
\epsffile{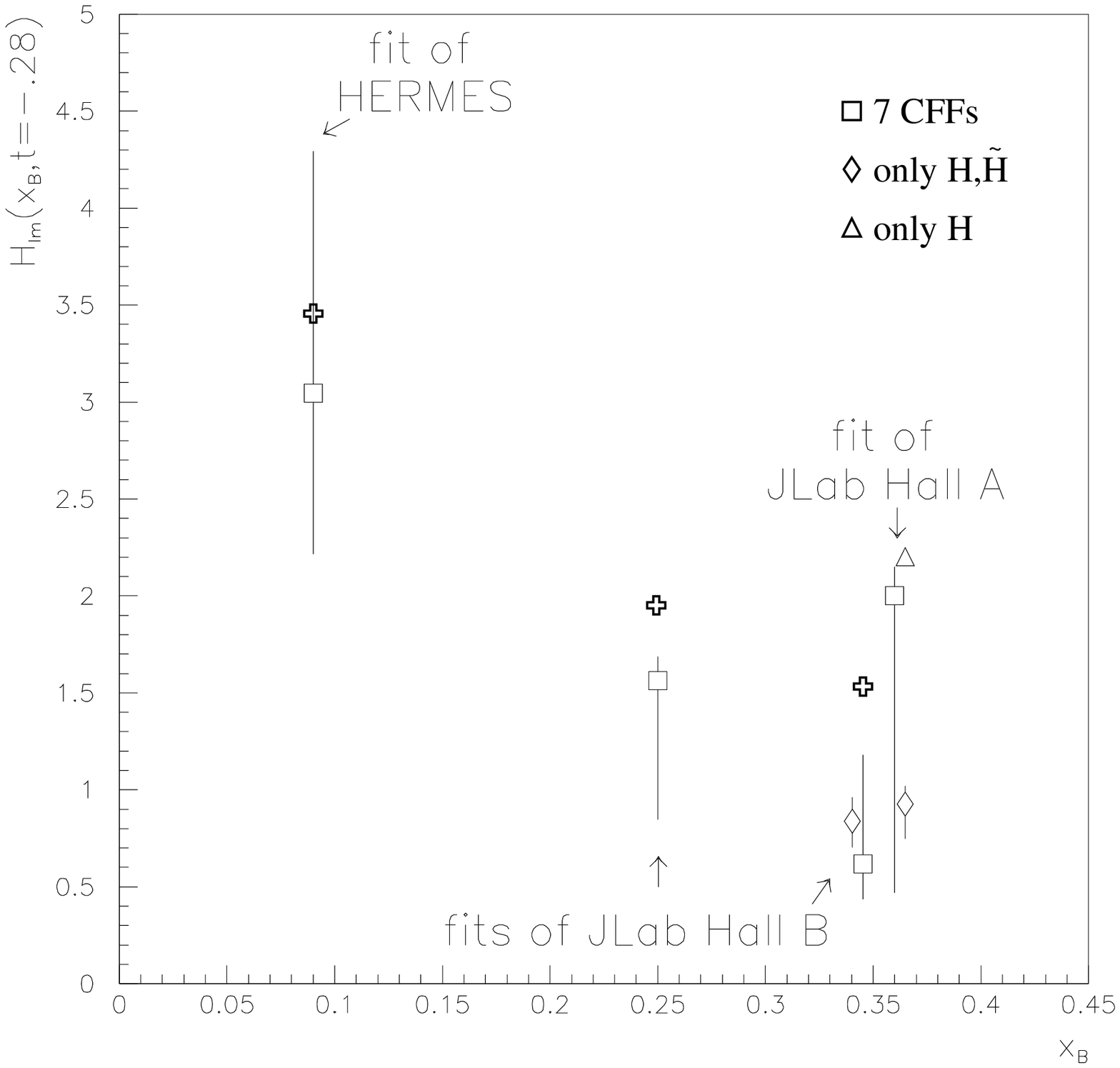}
\caption{$x_B$ dependence at fixed $-t$=0.28 GeV$^2$ of the fitted $H_{Im}$ 
(empty squares) according to our analyzes of the JLab Hall A 
data~\cite{fitmick} ($<x_B>$=0.36), of the HERMES data~\cite{fithermes} 
($<x_B>$=0.09) and of the present analysis ($<x_B>$=0.25 and $<x_B>\approx$0.35),
using the seven CFFs. The open diamond, slightly shifted left (for visibility) 
of the open square point at $x_B\approx$0.35 is the result of the CLAS BSAs
and lTSA at $x_B\approx$0.35 using only the $H$ and $\tilde{H}$ GPDs. 
The open triangle, slightly shifted right (for visibility) 
of the open square point at $x_B\approx$0.36 is the result of the JLab Hall A 
unpolarized and beam-polarized cross sections at $x_B$=0.36 using
only the $H$ GPD. The open diamond, slightly shifted right (for visibility) 
of the open square point at $x_B\approx$0.36 is the result of the JLab Hall A 
unpolarized and beam-polarized cross sections at $x_B$=0.36 using
only the $H$ GPD and $\tilde{H}$ GPDs. The empty cross indicates the VGG 
prediction.}
\label{fig:xdep}
\end{figure}

We recall that we were able in earlier work to also extract values 
for $H_{Im}$ at different $x_B$ values and at almost the $t$ values 
considered here ($\approx$-0.28 GeV$^2$). For recall, in Ref.~\cite{fitmick}, 
we fitted the JLab Hall A data which
have $<x_B>\approx$0.36 and in Ref~\cite{fithermes}, we fitted the HERMES data
which have $<x_B>\approx$ 0.09. While the JLab Hall A data were
taken precisely at $<-t>$=0.28 GeV$^2$, the HERMES data were given
for $<-t>$=0.20 GeV$^2$ and $<-t>$=0.42 GeV$^2$. In a very simplistic way,
we decide to interpolate between these two $-t$ values by simply
averaging our fitted $H_{Im}$ CFFs at these two $-t$ values. We also average 
quadratically the positive and negative error bars. We thus end up with
some average HERMES $H_{Im}$ CFF at $<-t>\approx$ 0.30 GeV$^2$
and $<x_B>\approx$ 0.09.
We can then obtain a $x_B$ dependence of our fitted $H_{Im}$'s using
our JLab and HERMES analysis results. Fig.~\ref{fig:xdep} shows this $x_B$ 
dependence, compiling our results from the independent analysis of the 
JLab Hall A data, the HERMES data and the presently analyzed CLAS data. 

It turns out that around $x_B$=0.35, both JLab Hall A and CLAS 
data are available. It is comforting to note the decent
agreement, within error bars, of the two extracted values for $H_{Im}$. 
True, the error bars are not small and it might appear not so 
challenging to have an agreement with such uncertainties. 
Nevertheless, we would rather take this aspect as support for our 
realistic evaluation of the error bars on our fitted CFFs.
 
In order to illustrate this point, we also plot 
in Fig.~\ref{fig:xdep} (open diamond around $x_B$=0.35) 
our fit results of the CLAS BSAs and lTSA 
at $x_B$=0.35 with only $H$ and $\tilde{H}$ as fitting parameters, i.e. 
four CFFs instead of seven. It is seen that the central value result which,
as could be expected, has a quite smaller error bar, is slightly shifted
with respect to the central value result when the seven CFFs are
taken into account. However, both results remains well compatible
within error bars.
We did the same exercice with the JLab Hall A data at $x_B$=0.36.
In Fig.~\ref{fig:xdep}, on the one hand, the open triangle shows the result of the 
fit to the JLab Hall A unpolarized and beam-polarized cross sections at 
$<-t>\approx$ 0.28 GeV$^2$, using only the $H$ GPD, i.e. two CFFs.
On the other hand, the open diamond around $x_B$=0.36 the same fit using only 
$H$ and $\tilde{H}$, i.e. four CFFs. Let us mention that these two fits give 
$\chi^2$ values very close to one and describe very well the 
data points.
Since these two fits use only a few CFFs, the error bars on 
the resulting $H_{Im}$'s are quite (artificially ?) small compared to the 
error bar resulting from the fit with seven CFFs (open square). However, it 
is striking to see how the central value of $H_{Im}$ can shift. With only 
$H$ entering the JLab Hall A data fit, $H_{Im}$ (open triangle
near $x_B$=0.36) is around 
two, at the top of the error bar of the seven CFFs fit result (open square
near $x_B$=0.36). This slight increase of $H_{Im}$ is actually very consistent 
with the result of Ref.~\cite{rv}
which used the same assumption of $H$ dominance. However, this
result is now clearly inconsistent with the result for  $H_{Im}$
issued from our fit of the CLAS data (open square near $x_B$=0.35). 

Turning to the configuration where both $H$ and $\tilde{H}$ enter 
the fit of the same JLab Hall A data, $H_{Im}$ has now dramatically dropped by a 
factor more than two (open diamond near $x_B$=0.36). However, it has now actually 
become compatible with the result 
issued from the CLAS data, either open square or diamond near $x_B$=0.35. 
Indeed, note that both CLAS $H_{Im}$ results, i.e. issued
from the fit with only $H$ on the one hand and with the seven CFFs
on the other hand, are compatible within error bar. This is a clear-cut illustration of the 
meaning of the large error bars which result from our fits when we use seven 
CFFs: they cautiously and realistically reflect the underconstrained nature 
of our problem, i.e. fitting only a couple of observables with many 
parameters (i.e. the CFFs), and all the variation and potential weight of 
these parameters.

To summarize this discussion, in the framework
of our analysis (i.e. leading twist and leading order QCD and
the few kinematic approximations mentionned earlier), it doesn't seem 
to us possible to find a consistent value of $H_{Im}$ to fit both the 
JLab Hall A and CLAS data if only $H$ enters the fit. The minimum 
scenario seems that $\tilde{H}$ be included, the ultimate one being of course 
that all CFFs be included. It is interesting to mention that Ref.~\cite{kumer}
reached some similar conclusion in a model-dependent approach, confirming 
the hint that GPDs other than $H$ (and possibly $\tilde{H}$) 
do play a significant role at the JLab kinematics.
Let us stress again that at this kinematic point, $<x_B>\approx$0.35, 
our values of $H_{Im}$ were determined by the fitting of independent
DVCS experiments, i.e. JLab Hall A and CLAS, and rather different observables:
polarized and unpolarized cross sections for the JLab Hall A analysis
and BSA and lTSA for the present CLAS analysis.
Although beam polarized observables are common to the two
experiments, it is encouraging to observe that different paths
can lead to consistent results, as it should be.

Now, more generally, taking into account the HERMES data, we observe 
in Fig.~\ref{fig:xdep} that 
the general tendency is that, at fixed $t$, $H_{Im}$ increases with decreasing 
$x_B$. This is reminiscent of the $x_B$ dependence of standard
parton distributions. The VGG prediction is also shown in Fig.~\ref{fig:xdep}
and, although it overestimates most of the fitted central
values, it displays the same behavior.

\begin{figure}[htb]
\epsfxsize=9.cm
\epsfysize=10.cm
\epsffile{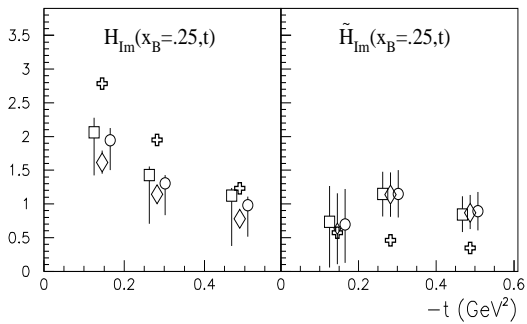}
\vspace{-4.2cm}
\caption{$t$-dependence of our fitted $H_{Im}$ and $\tilde{H}_{Im}$
CFFs at the kinematic points of Table~\ref{tab:tdep}. The empty squares 
(circles) show our results when the boundary values of the domain over which 
the CFFs are allowed to vary is 5 (3) times the VGG reference values. 
The empty diamonds indicate the results of our fits when only the $H$ and 
$\tilde{H}$ GPDs are taken as fitting parameters. The 
empty crosses indicate the VGG prediction. At the lowest $t$ value
of $\tilde{H}_{Im}$, the empty cross and diamond happen to be superimposed.}
\label{fig:tdep}
\end{figure}

We now finally turn our attention to the $t$ dependence of the lTSAs. The
CLAS collaboration has extracted the lTSAs at fixed $x_B$ ($\approx$0.25) 
for three different $<-t>$ values: 0.15, 0.24 and 0.43 GeV$^2$.
For each of these $-t$ points, we can identify three BSAs which have 
approximately the same $x_B$ and $-t$ values with, however, different
$Q^2$ values. We list those points in table~\ref{tab:tdep}. Comforted by
our reasonable results presented in Fig.~\ref{fig:res}, we adopt
the same approach and fit, with the seven CFFs as fitting parameters,  
simultaneously the three BSAs and the lTSA at each of the 
three ($x_B$, $-t$) points of Table~\ref{tab:tdep}, which all
have a common $x_B$ value (i.e. $\approx$ 0.25). Again, only the $H_{Im}$ 
and $\tilde{H}_{Im}$ CFFs systematically come out from our
fits with finite error bars. 
Fig.~\ref{fig:tdep} shows our results and reveals the $t$-dependence
(at $x_B\approx$0.25) of the $H_{Im}$ and $\tilde{H}_{Im}$ CFFs. We again
display for each $t$ value two results corresponding, like in
Fig.~\ref{fig:res}, to different boundary values for the domain of
variation allowed for the CFFs, i.e. $\pm$5 (empty squares) and $\pm$3 (empty
circles) times the VGG reference values. We also show 
in this figure the VGG predictions (empty crosses).
 
\begin{table}[h]	
\begin{tabular}{||c||c|c|c||}
\hline \hline
     & $<x_B>$& $<-t>$& $<Q^2>$ \\ \hline
lTSA & 0.27 & 0.15 & 1.82 \\ \hline
BSA1  & 0.24 & 0.15 & 1.65 \\ \hline
BSA2  & 0.24 & 0.14 & 1.89 \\ \hline
BSA3  & 0.25 & 0.14 & 2.16 \\ \hline\hline\hline
lTSA & 0.27 & 0.24 & 1.82 \\ \hline
BSA1  & 0.25 & 0.28 & 1.69 \\ \hline
BSA2  & 0.25 & 0.28 & 1.95 \\ \hline
BSA3  & 0.25 & 0.28 & 2.21 \\ \hline\hline\hline
lTSA & 0.27 & 0.43 & 1.82 \\ \hline
BSA1  & 0.25 & 0.49 & 1.70 \\ \hline
BSA2  & 0.25 & 0.49 & 1.95 \\ \hline
BSA3  & 0.25 & 0.49 & 2.20 \\ \hline\hline

\hline
\end{tabular}	
\caption{Selection, for our $t$-dependence study, of the three ($x_B$, $-t$)
kinematic points measured by the CLAS collaboration which have 
1 lTSA and 3 BSAs at approximately the same $x_B$ values ($\approx$ 0.25).} 
\label{tab:tdep}
\end{table}	

In Fig.~\ref{fig:tdep}, regarding $H_{Im}$, we note a smooth and typical 
fall-off with $-t$ which was also observed in our previous JLab Hall A 
and HERMES studies~\cite{fitmick,fithermes}. The figure also confirms that
the standard VGG parametrisation, in general, overestimates our fitted 
values. This is particularly the case at low $t$ (this was also
observed at HERMES energies~\cite{fithermes}). Regarding $\tilde{H}_{Im}$,
although the uncertainties are large, the $t$-slope appears
to be much less pronounced and it even seems that there is a 
drop towards 0 as $t$ goes to 0 (although a constant and flat $t$-dependence
can also be in order within error bars). We find again that VGG underestimates
this CFF, in particular as $\mid t\mid$ grows. Overall, the VGG $t$-slope
is markedly different from the one of the fitted $\tilde{H}_{Im}$. 

Finally, the diamonds in 
Fig.~\ref{fig:tdep} show the results of our fit when only the $H$ and 
$\tilde{H}$ GPDs are taken as fitting parameters, i.e. setting to 0
all other GPDs. We observe the same features as previously
(see Fig.~\ref{fig:res}). The central values are in very good agreement 
with the ones determined from the fits in which all
CFFs were included (though with a systematic
decrease by $\approx$ of 15\% of the central value of $H_{Im}$).
The main effect
is to reduce the uncertainties on the fitted CFFs: very strongly 
for $H_{Im}$ and only slightly for $\tilde{H}_{Im}$.

To summarize this work, we have analyzed the beam spin asymmetry and the longitudinally 
polarized target spin asymmetry of the Deep Virtual Compton Scattering 
process recently measured by the CLAS collaboration. We have used
a fitter code, largely model-independent, based on the QCD leading-twist
and leading order DVCS+BH amplitude, which takes as fitting parameters GPD CFFs. 
Even though we fit only two asymmetry observables with seven CFFs,
two CFFs, $H_{Im}$ and $\tilde{H}_{Im}$, come out systematically 
from our fits with stable and well defined central values and uncertainties
(of the order of 30\% in average). The reason is that the two observables we fit are well 
known to be dominantly sensitive to these two CFFs. It is worth
noting that with only BSAs to fit, there is no convergence of our fits,
while with the addition of a single observable, i.e. the lTSA, solutions
become relatively well defined.

In this work, a few approximations have been made, mostly due to the present lack
of sufficiently precise and numerous data. We recall that only
six lTSAs were available to us: for three $-t$ values at fixed $x_B$ and
$Q^2$ and for three $x_B$ values at fixed $-t$ and $Q^2$ values.
The approximations that we did were to simultaneously fit BSAs and lTSAs
taken at slightly different $x_B$ and $-t$ values and also at different $Q^2$ values.
This latter approximation is, in any case, along the line of the main starting
assumption of this work: the dominance of the QCD leading twist 
and leading order of the DVCS amplitude. In those conditions, we have been able to 
determine the $x_B$- and $t$- dependences of the $H_{Im}$ and $\tilde{H}_{Im}$ 
CFFs (respectively at fixed $t$ and fixed $x_B$). In particular, we put in evidence a 
much flatter $t$-dependence for $\tilde{H}_{Im}$ than for $H_{Im}$. 
We also illustrated, by comparing our fits at roughly the same
kinematics of the independent JLab Hall A and CLAS data, the importance 
and, even the necessity, of taking into account several GPDs in order to 
obtain compatible results.

While there have lately been a couple of other works aiming at fitting DVCS data and 
extracting $H_{Im}$, this is the first one allowing access to $\tilde{H}_{Im}$ and, in 
a largely model-independent way, determine some first numerical value for it.
The ``price" to pay for our model-independency is that we obtain relatively large 
uncertainties.
Several DVCS experiments aiming at measuring more precisely the observables analyzed 
in this work and also aimed at measuring new observables, such as transversely
polarized target spin asymmetries and cross sections are under way in the near
future. We expect our fitting techniques to be more and more fruitful
and efficient as these new precise and numerous data become available, along
with theoretical GPD modelling progress which can reduce the
domain of variation of the fitted CFFs or their number.

\end{document}